\documentclass[preprint]{elsarticle}
    \newcommand{\be}[1]{\begin{equation}\label{#1}}
    \newcommand{\ba}[1]{\begin{eqnarray}\label{#1}}

    \newcommand{\rd}{{\rm d}}
    
    \newcommand{\pa}[1]{\left(#1\right)}
    \newcommand{\paq}[1]{\left[#1\right]}
    \newcommand{\M}{{\rm M_{\rm P}}}
    \def\ee{\end{equation}}
    \def\ea{\end{eqnarray}}

\usepackage{amssymb}
\usepackage{graphicx}
\usepackage{graphicx,epsf}
\usepackage{epsfig}
\journal{Physics Letters B}
\begin{document}
\begin{frontmatter}

\title{Dynamical Dark Energy and Spontaneously Generated Gravity}
\author{Alexander Y. Kamenshchik\corref{cor1}}
\address{Dipartimento di Fisica and INFN, Via Irnerio 46,40126 Bologna,
Italy\\
L.D. Landau Institute for Theoretical Physics of the Russian
Academy of Sciences, Kosygin str. 2, 119334 Moscow, Russia}\ead
{Alexander.Kamenshchik@bo.infn.it}
\author{Alessandro Tronconi}
\address{Dipartimento di Fisica and INFN, Via Irnerio 46,40126 Bologna,
Italy}\ead{Alessandro.Tronconi@bo.infn.it}
\author{Giovanni Venturi}
\address{Dipartimento di Fisica and INFN, Via Irnerio 46,40126 Bologna,
Italy}\ead{Giovanni.Venturi@bo.infn.it}

\begin{abstract}
We study the cosmological evolution of an induced gravity model with a scale symmetry breaking potential for the scalar field. The radiation to matter transition, following inflation and reheating, influences the dynamics of such a field through its non minimal coupling. We illustrate how, under certain conditions on the potential, such a dynamics can lead to a suitable amount of dark energy explaining the present accelerated expansion.
\end{abstract}

\end{frontmatter}
\section{Introduction}
In induced \cite{sakharov}, or spontaneously generated, gravity the gravitational coupling (constant) and interaction arise as a quantum effect, in particular as a one-loop effect in some fundamental interaction, or through spontaneous symmetry breaking, always in association with the coupling of the curvature scalar to some hitherto unknown scalar field. Thus gravity itself would not be associated with ``fundamental physics'' but would be an emergent effect in which the conventional formulation is the low energy limit.\\
Induced gravity models have been applied to cosmology for several years, beginning with the original model for a time varying gravitational coupling in the presence of matter suggested by Brans and Dicke \cite{BD}
\footnote{Indeed the technique of the reconstruction of the scalar field potential providing a given cosmological evolution, well known for the case of the minimally coupled scalar field, was generalized for the case of induced gravity in paper \cite{ind-rec}.}.
Subsequently a simple scalar field model for induced gravity which avoided the excessive time variation of the gravitation coupling was introduced \cite{CV}. This latter model was globally scale invariant (that is did not contain any dimensional parameter) and spontaneous symmetry breaking in such a context not only generated the gravitational constant but also a cosmological constant, corresponding to dark energy. The introduction of both radiation and matter further showed that the model led to Einstein gravity plus a cosmological constant as a stable attractor among homogeneous cosmologies and was therefore a viable dark energy model for a range of scalar field initial conditions \cite{induced}. One drawback of the above simple model is that it exhibits stability for any (constant) value of the
scalar field and the desired value must be determined either by the presence of a condensate or of quantum effects.\\
Subsequently inflation and reheating were studied in an induced gravity model having diverse (scale) symmetry breaking potentials \cite{Cerioni}. In particular the potentials examined were either associated with the presence of a condensate (Landau-Ginzburg) or quantum effects (Coleman-Weinberg). The latter case was studied both for an effective potential inspired by the flat space results and the potential obtained in a de Sitter background \cite{Tronconi}.
\\
For a potential explicitly exhibiting symmetry breaking, the gravitational constant is determined by the value of the scalar field at the end of inflation and reheating when the field is at the minimum of the potential. At this point, however, one no longer appears to have a cosmological constant (or dark energy).\\
The scope of this work is to illustrate how a dynamical (time dependent) cosmological constant could arise during the radiation-matter transition. This is of interest for the coincidence problem \cite{Zlatev:1998tr} insofar as the transition from radiation to matter will, through the presence of the coupling of the scalar field to the Ricci scalar, displace the scalar field from the potential minimum, thus generating a cosmological constant. Naturally the cosmological constant generated must be such as not to prevent the formation of collapsed structures and the return to the minimum be slow enough to respect the current limits on the time variation of the gravitational coupling. We shall begin by examining the case for a simple Landau-Ginzburg potential and then see how it must be modified to fully satisfy the constraints.\\
Let us note that besides induced gravity models, there are models wherein the term with non-minimal coupling between the scalar field and the scalar curvature coexists with the standard  Einstein-Hilbert term. In particular, the models wherein the inflaton 
scalar field was non-minimally coupled to gravity have been demonstrated to have some advantages with respect to the minimally coupled models 
\cite{inf-non}. The hypothesis, identifying  a non-minimally coupled inflaton field with Higgs boson \cite{BezShap} has allowed to establish relations between the observable data coming from cosmology with those from particle physics \cite{inf-non1}.  
While giving, in general, more flexibility in tuning the parameters of the model under consideration, the non-minimally coupled 
gravity in the presence of the Einstein-Hilbert term, loses in comparison with the induced gravity its minimality and an 
attractive feature such as a capacity to treat gravity as quantum effect. We shall point out later that for our model, the inclusion of the 
Hilbert-Einstein term does not give any advantages and so it makes sense to just consider induced gravity. 
\section{The model}
Consider the following homogeneous Lagrangian density for a general induced gravity (IG) model
\be{lag}
\mathcal{L}=\sqrt{-g}\paq{\frac{\gamma}{2}\sigma^{2}R+\frac{1}{2}\dot\sigma^{2}-V(\sigma)}
\ee
on a spatially flat RW background $ds^{2}=dt^{2}-a(t)^{2}d\vec x\cdot d\vec x$.\\
In the presence of barotropic cosmological fluids, the effective Friedmann and Klein-Gordon equations for such a model are
\be{freqIG}
H^{2}=\frac{1}{3\gamma\sigma^{2}}\pa{H^{2}\frac{\sigma'^{2}}{2}+V(\sigma)+\rho_{M}+\rho_{R}-6\gamma H^{2}\sigma\sigma'}
\ee
and
\be{kgeqIG}
H^{2}\paq{\pa{\sigma^{2}}''+\pa{3+\frac{\pa{H^{2}}'}{2H^{2}}}\pa{\sigma^{2}}'}=\frac{2}{1+6\gamma}\pa{4V-\sigma \frac{\rd V}{\rd \sigma}+\rho_{M}}
\ee
with the prime denoting a derivative w.r.t. the number of e-folds $N\equiv \ln\pa{a/a_{0}}$, $\rho'_{M}=-3\rho_{M}$ (dust) and $\rho_{R}'=-4 \rho_{R}$ (radiation).
We may re-write the equations (\ref{freqIG}) and (\ref{kgeqIG}) in terms o the variables $x\equiv\gamma \sigma^{2}$ and $y=H^{2}$ as
\be{FR}
y=\frac{V(x)+\rho_{M}+\rho_{R}}{3x\pa{1-\frac{1}{24\gamma}\frac{x'^{2}}{x^{2}}+\frac{x'}{x}}}
\ee
and
\be{KG}
y\paq{x''+\pa{3+\frac{y'}{2y}}x'}=\frac{2\gamma}{1+6\gamma}\pa{4V-2x\frac{\rd V}{\rd x}+\rho_{M}}
\ee
where the new variables $x$ and $y$ are positive definite.\\
The scalar field $\sigma$ is associated with the observed value of Newton's constant and a stabilizing potential $V(\sigma)$ in the form of a symmetry breaking potential is generally assumed in order to set its value to $\sigma_{0}$ in such a way that $\gamma \sigma_{0}^{2}=x_{0}=\M^{2}$. In the absence of the cosmological fluid, when the fields sits in the minimum of the potential, its v.e.v. is different from zero and its zero point energy is null. A natural choice for $V(\sigma)$ is that of a Landau-Ginzburg potential:
\be{LG}
V(\sigma)=\frac{\lambda}{4}\pa{\sigma^{2}-\sigma_{0}^{2}}^{2}=\frac{\lambda}{4\gamma^{2}}\pa{x-x_{0}}^{2}.
\ee 
In a previous paper \cite{induced} we studied an IG dark energy model with just a quartic potential ($\sigma_{0}=0$). Such a model was shown to be viable as a dark energy candidate due to its scale invariance and the presence of the de Sitter attractor in the absence of cosmological fluids. On setting $\sigma_{0}=0$ in (\ref{LG}) the model is recovered. These attractors are stable but there is no dynamical mechanism selecting among them and fitting observations requires some sort of fine tuning of the initial conditions. \\
\subsection{Absence of Cosmological Fluids}
Let us first consider the case $\rho_{M}=\rho_{R}=0$ and a generic potential $V(\sigma)$. It is clear from (\ref{FR},\ref{KG}) that a de Sitter solution ($y=\bar y={\rm const.}$) exists when 
\be{FRDS}
\bar y\pa{1-\frac{1}{24\gamma}\frac{x'^{2}}{x^{2}}+\frac{x'}{x}}=\frac{V}{3x}
\ee
and
\be{KGDS}
\frac{\bar y}{3}\pa{\frac{x''}{x}+3\frac{x'}{x}}=\frac{4\gamma}{1+6\gamma}\paq{\frac{V}{3x}-x\frac{\rd}{\rd x}\pa{\frac{V}{3x}}}.
\ee
The second equation can be re-written as
\be{KGDS2}
\frac{\bar y}{3}\;\paq{\frac{\rd}{\rd N}\pa{\frac{x'}{x}}+\pa{\frac{x'}{x}}^{2}+3\frac{x'}{x}}=\frac{4\gamma}{1+6\gamma}\paq{\frac{V}{3x}-\frac{x}{x'}\frac{\rd}{\rd N}\pa{\frac{V}{3x}}}
\ee
and finally takes the form
\ba{KGDS3}
&&\frac{\rd}{\rd N}\pa{\frac{x'}{x}}+\pa{\frac{x'}{x}}^{2}+3\frac{x'}{x}\nonumber\\
&&=\frac{12\gamma}{1+6\gamma}\paq{\pa{1-\frac{1}{24\gamma}\frac{x'^{2}}{x^{2}}+\frac{x'}{x}}+\frac{x}{x'}\frac{\rd}{\rd N}\pa{\frac{1}{24\gamma}\frac{x'^{2}}{x^{2}}-\frac{x'}{x}}}.
\ea
Eq. (\ref{KGDS3}) can be solved for $x'/x=\bar \delta={\rm const}$ with $\bar \delta$ satisfying the algebraical equation
\be{KGDSA}
\bar \delta^{2}+3\bar \delta=\frac{12\gamma}{1+6\gamma}\paq{\pa{1-\frac{1}{24\gamma}\bar \delta^{2}+\bar \delta}}.
\ee
The solutions of Eq. (\ref{KGDSA}) are $\bar \delta_{1,2}=-2,\frac{4\gamma}{1+4\gamma}$. On substituting such solutions into the original equations (\ref{FRDS}) and (\ref{KGDS}) one easily verifies that $\bar\delta_{2}$ is associated with the potential $V\propto x\propto \sigma^{2}$. The solution $\delta_{1}$, on the other hand, actually corresponds to the trivial case $\bar y=0$ and $V=0$ (see \cite{ind-rec} for details).\\
The case $\bar \delta=0$ should be studied separately. In this last case $x=\bar x$ and Eq. (\ref{KGDS}) becomes
\be{KGDS0}
\left.\frac{V}{3x}-x\frac{\rd}{\rd x}\pa{\frac{V}{3x}}\right|_{x=\bar x}=0
\ee
which is identically satisfied when $V\propto x^{2}\propto \sigma^{4}$, and for more general potentials it constrains $\bar x$. For the potential (\ref{LG}) this last equation leads to:
\be{KGDSLG}
\frac{\lambda}{12\gamma^{2}}\frac{\bar x^{2}-x_{0}^{2}}{\bar x}=0\Rightarrow \bar x=x_{0}.
\ee
\subsection{Matter Domination}
The energy density of the matter fluid enters in the evolution equation of the scalar field as an external force which increases the expectation value of the scalar field (see Eq. (\ref{KG})). In the presence of a potential of the form given by (\ref{LG}) one must also take into account the effect of such a potential on the r.h.s. of Eq. (\ref{KG}) ``pulling'' the field toward its attractor $x_{0}$:
\be{poteffect}
4V-2x\frac{\rd V}{\rd x}=\frac{\lambda}{\gamma^{2}}x_{0}\pa{x-x_{0}}\Rightarrow V=\frac{\gamma^{2}}{\lambda\,x_{0}^{2}}\pa{2V-x\frac{\rd V}{\rd x}}^{2}.
\ee 
If the scalar field is sitting on the minimum of (\ref{LG}) at the onset of matter domination, its dynamics can be described by two phases.
During the first phase the field increases because the``pulling'' force is negligible at the minimum of the potential ($V(x_{0})=0$, $dV/dx(x_{0})=0$) w.r.t. the ``pushing'' effect due to matter, i.e. $\rho_{M}\gg V$. One can then approximate Eqs. (\ref{FR}), (\ref{KG}) by neglecting $V$ and its derivatives. During the second phase the energy density of the scalar field dominates over matter and the field rolls down the potential and finally ends up oscillating with a decaying amplitude around the minimum $x_{0}$. If the scalar field rolls down slowly enough then it behaves as dark energy and drives cosmic acceleration.\\
At the onset of matter domination the Friedmann equation 
\be{FRp1}
y\simeq\frac{\rho_{M}}{3x\pa{1-\frac{1}{24\gamma}\frac{x'^{2}}{x^{2}}+\frac{x'}{x}}}\Rightarrow \rho_{M}\simeq 3xy\pa{1-\frac{1}{24\gamma}\frac{x'^{2}}{x^{2}}+\frac{x'}{x}}
\ee
well determine the approximate dynamics. The Klein-Gordon equation can be cast into the following form
\ba{KGp1}
y\paq{x''+\pa{3+\frac{y'}{2y}}x'}&=&\frac{2\gamma}{1+6\gamma}\rho_{M}\Rightarrow\nonumber\\
\Rightarrow \paq{\frac{x''}{x}+\pa{3+\frac{y'}{2y}}\frac{x'}{x}}&\simeq&\frac{6\gamma}{1+6\gamma}\pa{1-\frac{1}{24\gamma}\frac{x'^{2}}{x^{2}}+\frac{x'}{x}}
\ea
where $y'/y$ can be obtained by deriving (\ref{FRp1}) and using the continuity equation for $\rho_{M}$. One obtains
\be{ypoy}
\frac{y'}{y}=-3-\frac{x'}{x}+\frac{\frac{1}{24\gamma}\frac{\rd}{\rd N}\frac{x'^{2}}{x^{2}}-\frac{\rd}{\rd N}\frac{x'}{x}}{1-\frac{1}{24\gamma}\frac{x'^{2}}{x^{2}}+\frac{x'}{x}}
\ee
and finally Eq. (\ref{KGp1}) takes the following form:
\ba{KGp12}
&&\paq{\frac{\rd}{\rd N}\pa{\frac{x'}{x}}+\frac{1}{2}\pa{\frac{x'}{x}}^{2}+\pa{\frac{3}{2}+\frac{1}{2}\frac{\frac{1}{24\gamma}\frac{\rd}{\rd N}\frac{x'^{2}}{x^{2}}-\frac{\rd}{\rd N}\frac{x'}{x}}{1-\frac{1}{24\gamma}\frac{x'^{2}}{x^{2}}+\frac{x'}{x}}}\frac{x'}{x}}\nonumber\\
&&\simeq\frac{6\gamma\,}{1+6\gamma}\pa{1-\frac{1}{24\gamma}\frac{x'^{2}}{x^{2}}+\frac{x'}{x}}.
\ea
The above equation is a first order differential equation for the function $x'/x$ and has attractor solutions which can be found by solving Eq. (\ref{KGp12}) algebraically when $\frac{\rd}{\rd N}\pa{\frac{x'}{x}}=0$. One finds
\be{d12X}
\left.\frac{x'}{x}\right|_{1,2}=-2,\;\frac{4\gamma}{4\gamma+1}.
\ee
The solution $\frac{x'}{x}=-2$ was also found in the previous section, in the absence of fluids, and still corresponds to the trivial case $y=0$ and $\rho_{M}=0$.\\
Thus when matter dominates over the energy density of the scalar field, the field increases exponentially as
\be{expincrease}
x=x_{0}\exp\paq{{\frac{4\gamma}{1+4\gamma}\pa{N-N_{e}}}}
\ee
where $N_{e}$ is the value of $N$ at the matter-radiation equality (we take $N=0$ today), whereas the matter density decreases as $\rho_{M}=\rho_{M,0}\exp\pa{-3N}$ where $\rho_{M,0}$ is the energy density of matter today. This regime comes to an end when 
\be{balance}
\rho_{M}\simeq 4V-2x\frac{\rd V}{\rd x}\Rightarrow\frac{\rho_{M}}{2}\simeq\frac{\lambda}{2\gamma^{2}} x_{0}\pa{x-x_{0}}.
\ee
At the beginning of the second phase the above equation can then be recast in the following form:
\be{coinc1}
\rho_{M}\simeq \frac{4}{\epsilon_{x}} V\simeq \frac{4}{\epsilon_{x}} \rho_{\sigma}
\ee
with $\epsilon_{x}=(x-x_{0})/x_{0}$, $\rho_{\sigma}$ the energy density of the scalar field \cite{induced}, and the last equality in (\ref{coinc1}) holds approximately only when the scalar field begins to dominate and slowly varies in time. In particular if $\epsilon_{x}< 1$ one has $\rho_{\sigma}< \rho_{M}$ when the scalar field begins to roll toward the minimum of the potential. For small values of gamma the condition $\epsilon_{x}< 1$ is generally true and one may have a Universe dominated by dark energy today but not in the far past. The coincidence problems is then alleviated for such classes of dark energy models because reasonable conditions on $\gamma$ (small values of $\gamma$ are needed to satisfy solar system observational limits on the model) put a bound on the maximum value of $\rho_{\sigma}$.\\
On substituting the approximate evolution for $x(N)$ one gets:
\be{bal2}
\frac{\rho_{M,0}}{x_{0}^{2}}\exp\pa{-3N_{*}}\simeq\frac{\lambda}{\gamma^{2}} \paq{\exp\pa{\frac{4\gamma }{1+4\gamma}\pa{N_{*}-N_{e}}}-1}
\ee
where $N_{*}$ is the number of e-folds between the epoch when the scalar field begins to dominate over matter and today.\\
When the scalar field dominates, the evolution can be approximated by the solutions of Eqs. (\ref{FR},\ref{KG}) in the absence of fluids. Such solutions have been obtained in \cite{Cerioni} for the inflationary dynamics and are
\be{delsf}
\frac{x'}{x}\simeq\frac{4\gamma\pa{2-n}}{1+2\gamma\pa{n+1}}
\ee 
and
\be{epssf}
\frac{y'}{y}\simeq\frac{4\gamma\pa{2-n}\pa{n-1}}{1+2\gamma\pa{n+1}}
\ee 
where $n=\frac{\rd \ln V}{\rd \ln x}$. When $x'/x\ll 1$ and $y'/y\ll 1$ expressions (\ref{delsf}), (\ref{epssf}) are very accurate and describe a slow-roll dynamics similar to the inflationary case.

\section{Comparison with observations}
IG models as alternative theories of gravity are constrained by many observations.
The expectation value of the scalar field is associated with the ``fundamental'' Newton's constant $G_{N}$, namely the coupling which multiplies the Ricci scalar in the classical action. Furthermore the scalar field fluctuations around its homogeneous/classical trajectory couple to the trace of the energy-momentum tensor and they may modify the value of Newton's constant measured in Cavendish-like experiments. In particular if these fluctuations are massive they mediate short range interactions and the relation $x=\pa{8\pi G_{N}}^{-1}$ still holds at distances larger than their Compton wavelength. On the other hand, massless fluctuations of the scalar field, would generate a long range interaction and correspondently modify the measured Newton's constant by some factor $g(\gamma)$ with $\lim_{\gamma\rightarrow 0}g(\gamma)=1$. In the latter case the effective Newton's constant measured in Cavendish like experiment is$G_{N}^{\rm eff}=\pa{8\pi x}^{-1}g(\gamma)$.\\ 
The fluctuations of the scalar field around the minimum of the LG potential are indeed massive and their mass is $m^{2}=2\lambda x_{0}$. We thus expect modifications to General Relativity up to distances of the order $r\simeq \pa{\sqrt{\lambda x_{0}}}^{-1}$. Such modifications must be compatible with present day constraints on $G_{N}$. Newton's law of gravitation has been extensively tested at millimeter scales, at geophysical scales ($\simeq 10^{2}m$) and at astrophysical scales ($\simeq 10^{8}m$). The first two measurements succeeded in determining the value of the Newton's constant today with a precision of about $0.012\%$. Today we have $\pa{8\pi G_{N}}^{-1}=\M^{2}$ and $\M\simeq 2.436\cdot 10^{18}\;{\rm GeV}$.\\
In IG theories the ``fundamental'' Newton's constant also depends on time. At the cosmological level such a dependence plays a crucial role in the expansion of the homogeneous Universe. Given the matter content of the Universe at a certain time, larger values of $G_{N}$ may determine a faster expansion of the Universe compared to GR. The Big Bang Nucleosynthesis (BBN) era provides quite severe constraints on the expansion rate of the Universe in order to reproduce the correct abundances of light elements. Milder constraints are also imposed by CMB observations. In this latter case such constraints also depend on other cosmological parameters and are more cumbersome to impose. Stringent bounds on the time variation of $G_{N}$ today comes from solar system observations.  In particular the lunar laser ranging experiments impose the following limit on the time variation of Newton's constant: 
\be{dGbound}
\left|H^{-1}\dot G_{N}/G_{N}\right|_{\rm today}<0.02.
\ee
Io order to simplify the analysis of the viability of our dark energy model we set the minimum of LG potential at $x_{0}=\M^{2}$. When the scalar field sits in such a minimum, deep in the radiation domination era, the value of the Newton's constant is that of GR. The bounds imposed on $G_{N}$ by BBN are thus automatically satisfied. 
We first focus our analysis on the global evolution of the Universe from radiation domination until now. We assume a spatially flat Universe filled with radiation, matter and the homogeneous scalar field. In particular we fix the ratio between radiation and dust energy density today according to the best fit of the $\Lambda CDM$ model, which will be taken, from here on, as our fiducial model. These densities can be independently measured and we take
\be{rm/rr}
r'\equiv\frac{\rho_{R,0}}{\rho_{M,0}}=\frac{1}{3520}
\ee
in order to reproduce the standard transition between radiation and matter domination. This era is crucial in our simulations because at the onset of matter domination the scalar field begins to move from the bottom of the potential.\\
Given the above assumptions, the free parameters left in our model are $\gamma$, $\lambda$ and $\rho_{M,0}$. The energy density of the Dark energy component is defined as 
\be{rhoDE}
\rho_{DE}=3 M_{P}^{2}H^{2}-\rho_{M}-\rho_{R}
\ee
and today
\be{rhoDE0}
\rho_{DE,0}\simeq3 M_{P}^{2}H^{2}_{0}-\rho_{M,0}.
\ee

Let us note that, in principle, the  definition, given in Eq. (\ref{rhoDE0}), does not exclude a possibility of having a negative 
energy density for the dark energy. It does not occur in the range of the models parameters considered in the present paper.
However, the very opportunity of the change the sign of the effective density of the dark energy is interesting and 
and attractive from the point of view of some cosmological scenarios, considered, for example, in papers \cite{negative}.

In our simulation we replaced the quantities $\lambda$ and $\rho_{M,0}$ with $\beta$ and $\Lambda$ implicitly defined by
\be{defalpham}
\lambda\equiv \beta\Lambda,\;\rho_{M,0}\equiv\frac{\Lambda}{3\,\alpha_{m}}M_{P}^{4},
\ee
where $\alpha_{m}$ is related to the best $\Lambda CDM$ estimate of the ratio today between the dark energy density and matter density: 
\be{alpham}
\alpha_{m}\equiv \frac{1}{3}\left.\frac{\rho_{DE}}{\rho_{M,0}}\right|_{\Lambda CDM}\simeq 0.925
\ee
with $73\%$ of dark energy and $27\%$ of dust.
It turns out that the equation governing the dynamics of the scalar field is independent of $\Lambda$ and one only needs to fix $\gamma$ and $\beta$ to numerically solve for it. The evolution of the relative energy densities $\Omega_{M}$, $\Omega_{R}$ and $\Omega_{DE}$, with $\Omega_{i}\equiv\frac{\rho_{i}}{3 M_{P}^{2}H^{2}}$ are independent of $\Lambda$ as well. We then chose to fix $\gamma$ and $\beta$ by comparing the function
\be{weff}
w_{eff}\equiv-1+\frac{\rho_{M}+\frac{4}{3}\rho_{R}}{\rho_{M}+\rho_{R}+\rho_{DE}}
\ee
calculated for our model with that evaluated in the $\Lambda CDM$ model. In particular we minimized the distance between these two functions defined for $N\in\paq{-12,0}$ by assuming the following definition of distance between two generic real functions $f(N)$ and $g(N)$ over the interval $N\in\paq{N_{i},N_{f}}$:
\be{defdist}
d(f,g)\equiv\sqrt{\frac{\int_{N_{i}}^{N_{f}}dN\pa{f(N)-g(N)}^{2}}{N_{f}-N_{i}}}
\ee
and consequently defining the norm of $f(N)$ as $||f||\equiv d(f,0)$.
The numerical comparison gives $\gamma_{b}=0.014$ and $\beta_{b}=0.004$ and 
\be{distBF}
\frac{d(w_{eff}^{LG},w_{eff}^{\Lambda CDM})}{||w_{eff}^{\Lambda CDM}||}\simeq 0.08.
\ee
\begin{figure}[t!]
\centering
\epsfig{file=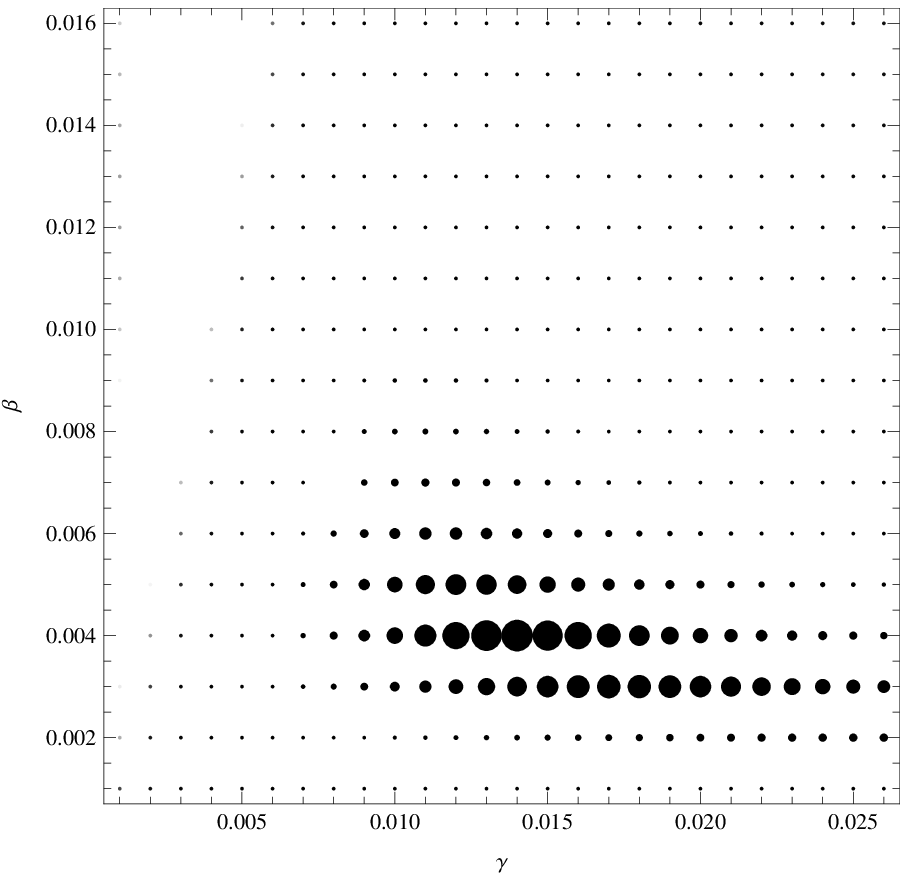, width=6 cm}
\epsfig{file=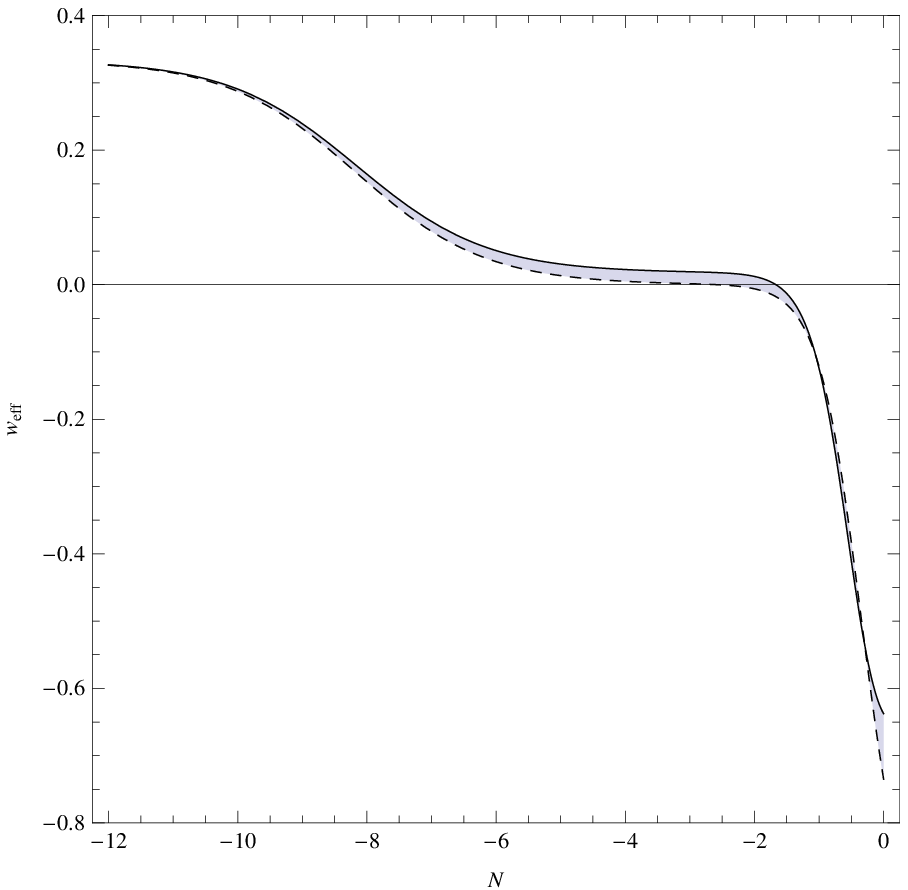, width=6 cm}
\caption{\it Results for the best fit in $\gamma$, $\beta$. In the figure on the left we plotted the distance as a function of the free parameters $\gamma$, $\beta$: larger dots correspond to smaller distances. In the figure on the right we plotted $w_{eff}$ calculated for the best fit $\gamma_{b}$, $\beta_{b}$ (solid line) and the same function in $\Lambda CDM$ (dashed line).}.
\label{fig1}
\end{figure}
The parameter $\Lambda$ can now be estimated, given $\gamma=\gamma_{b}$ and $\beta=\beta_{b}$, by minimizing the $\chi^{2}$ test fitting the supernovae Ia ``gold'' dataset (182 supernovae events). We obtained $\Lambda_{b}=8.3\times 10^{-121}$ and $\chi^{2}_{m}\simeq 165$ ($\#$ of degrees of freedom $=182-2-1=179$). In spite of the good agreement with the ``$w_{eff}$ test'' (\ref{distBF}) and supernovae observations the model is quite different from our fiducial model. In particular, on assuming a Cosmological Constant energy density of $\rho_{\Lambda}=9.6\times 10^{-121}\M^{4}$ the deviations of $H^{2}$ w.r.t. $\Lambda CDM$ evolution are close to $40\%$ today and the relative abundances of dark energy fluid and dust are about $61\%$ and $39\%$ respectively. Let us note that the above value of $\rho_{\Lambda}$ is that which minimizes the $\chi^{2}$ test for supernovae for the $\Lambda CDM$ model. On assuming the best estimate of $\rho_{\Lambda}$ ($\rho_{\Lambda}^{\rm best}\simeq9.04\times 10^{-121}\M^{4}$) the differences between our model and $\Lambda CDM$ are smaller but still quite large.\\
Serious problems arise if one calculates the value of the ``fundamental'' Newton's constant today and its variation in time: the maximum displacement of the scalar field from the minimum of the potential for $\gamma=\mathcal{O}(10^{-2})$ leads to a decrease of $G_{N}$ of about $30\%$. Moreover its time variation is about one order of magnitude larger than the present observational bounds. Furthermore one can calculate the mass of scalar fluctuations around the minimum of the potential and find the Compton wavelength of these fluctuations. It turns out that $\lambda_{\rm Compton}\simeq 10^{23}km$ and thus, for astrophysical scales, Newton's law should be modified by taking into account the effect of the fluctuations as well. The parameter $\gamma_{b}$ is then bounded by PPN constraints in modified gravity theories and is too large of about 5 orders of magnitude.\\
\begin{figure}[t!]
\centering
\epsfig{file=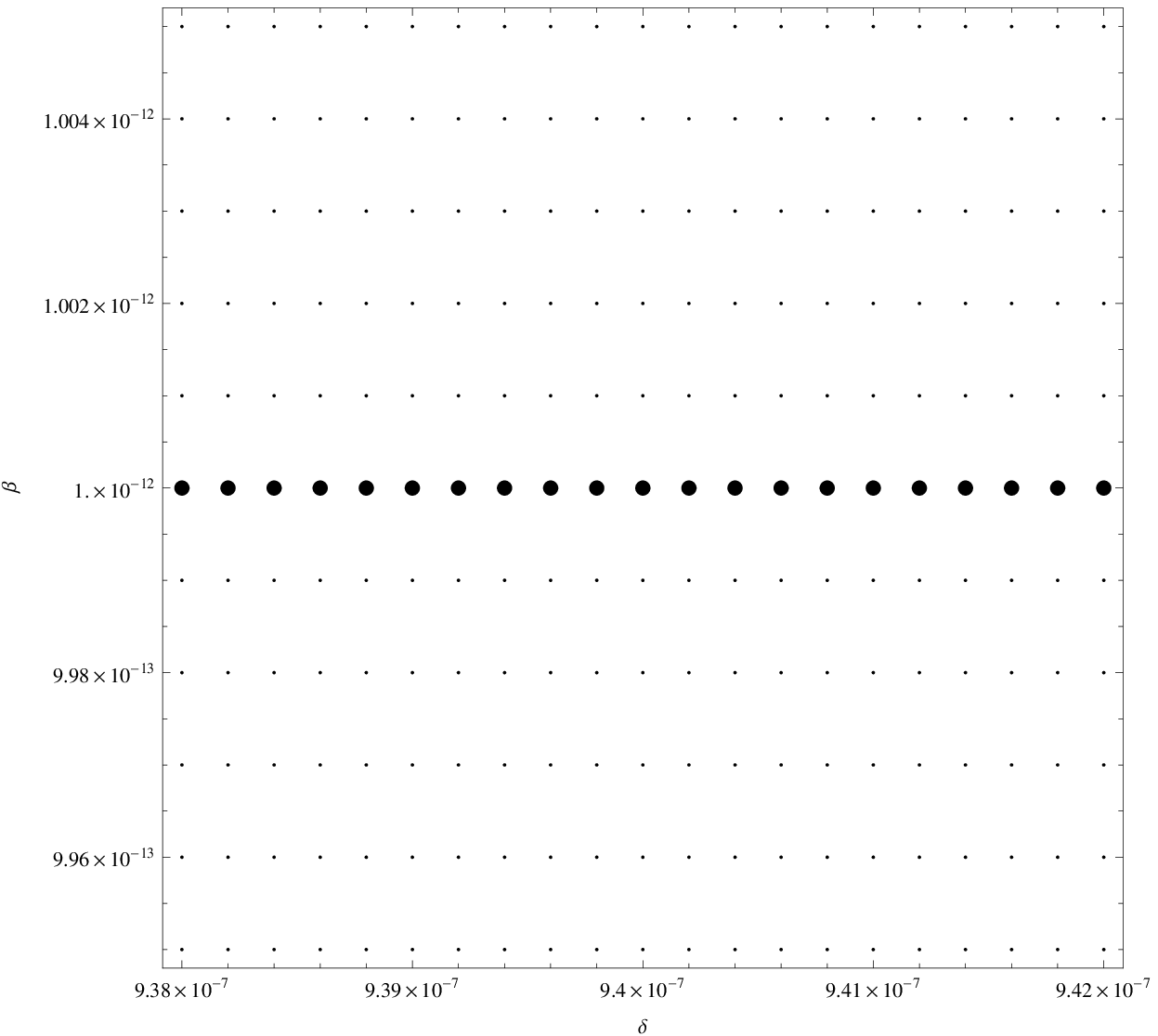, width=6 cm}
\epsfig{file=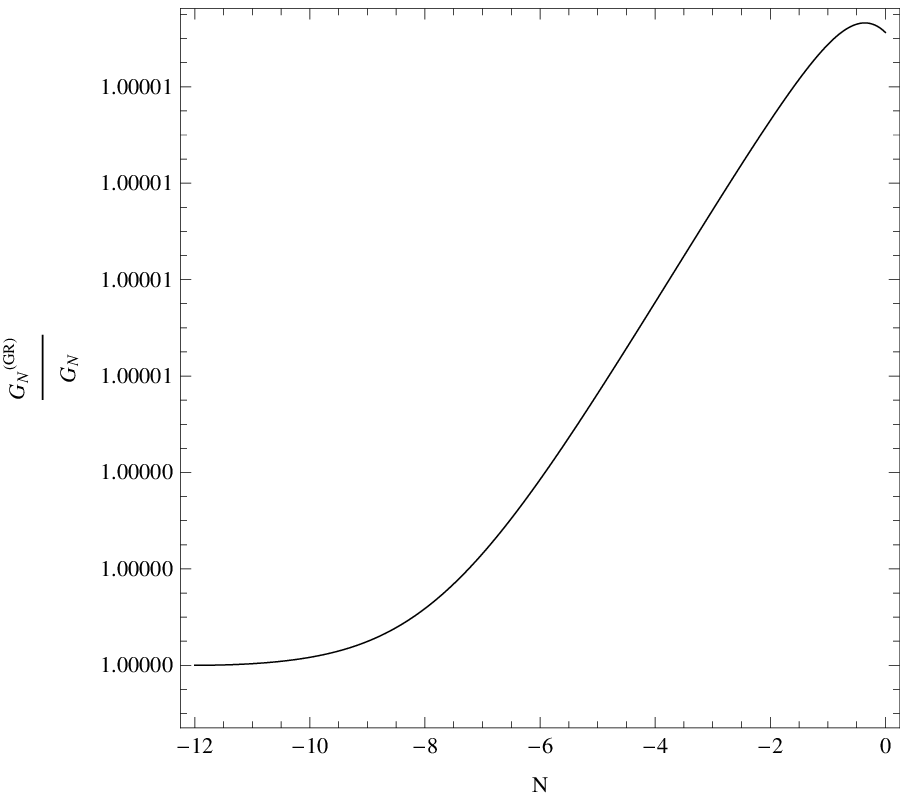, width=6 cm}
\caption{\it Results for the best fit in $\delta$, $\beta$ for the modified potential with $\gamma=5 \times 10^{-7}$. In the figure on the left we plotted the distance as a function of the free parameters $\delta$, $\beta$: larger dots correspond to smaller distances. In the figure on the right we plotted the ratio between the Newton's constant in General Relativity $G_{N}^{(GR)}$ and that in IG ($G_{N}\equiv (8\pi \gamma \sigma^{2})^{-1}$) calculated with $\delta_{b}$, $\beta_{b}$.}.
\label{fig2}
\end{figure}
In spite of the several deviations from observations it is worth emphasizing that our original, very simple, model with a LG potential is not many orders of magnitudes away from experimental data. One may thus expect that with small modifications of the shape of the original potential one can actually fit all the observations while preserving all the relevant features of the former potential.\\
A quite simple modification to the ``large field regime'' ($\sigma>\sigma_{0}$) dynamics could be done by multiplying each $\sigma_{0}$ term in the LG potential by some decreasing function of $M(\sigma)$ such as $M(\sigma_{0})=1$. On expanding the new modified potential $V_{M}(\sigma)$ around $\sigma_{0}$ one then gets 
\ba{exppot}
V_{M}&=&\frac{\lambda}{4}\pa{\sigma^{2}-\sigma_{0}^{2}M(\sigma)^{2}}^{2}\simeq \frac{\lambda}{4}\paq{\sigma_{0}^{2}+2\sigma_{0}\delta \sigma-\sigma_{0}^{2}\pa{1+2\frac{dM}{d\sigma}\delta\sigma}}^{2}\nonumber\\
&=&\frac{\lambda}{2}\sigma_{0}^{2}\pa{1-\sigma_{0}\frac{dM}{d\sigma}}\delta\sigma^{2}
\ea
where $\frac{dM}{d\sigma}$ is evaluated in $\sigma=\sigma_{0}$ and is negative. Thus the modified potential still has an absolute minimum at $\sigma_{0}$ and $V_{M}(\sigma_{0})=0$ as for the LG potential. Our aim is to obtain a potential which is closer to $\lambda \sigma^{4}$ in the large field regime and we are uninterested in the small field region.\\
We chose, for example
\be{Msigma}
M(\sigma)=e^{-\frac{\sigma-\sigma_{0}}{\delta\;\sigma_{0}}}
\ee
where $\delta$ is a free dimensionless parameter. 
Let us note that for small and even negative $\sigma$ the modification given by (\ref{Msigma}) gives particular shapes to the potential which depend on the value of $\delta$. 
In figure (\ref{fig4}) we plotted the ratio
\be{ratioV}
\epsilon(V)\equiv\sqrt{\frac{\pa{V-V_{0}}^{2}}{V_{0}^{2}}}
\ee
where $V$ is either the LG potential or $V_{M}$, with $M$ given by (\ref{Msigma}) and different choices of $\delta$, and the quartic potential $V_{0}\equiv \frac{\lambda}{4}\sigma^{4}$. The last potential produces a de Sitter expansion as the attractor of the dynamics and the ratio (\ref{ratioV}), expressed as a function of $\delta \sigma\equiv \frac{\sigma}{\sigma_{0}}-1$, measures the difference between some $V$ and the quartic potential itself at a given distance $\delta \sigma$ from the minimum $\sigma_{0}$ of $V$. The effect of $M$ is that of decreasing the ratio $\epsilon(V)$ for a given displacement $\delta \sigma$.\\
\begin{figure}[t!]
\centering
\epsfig{file=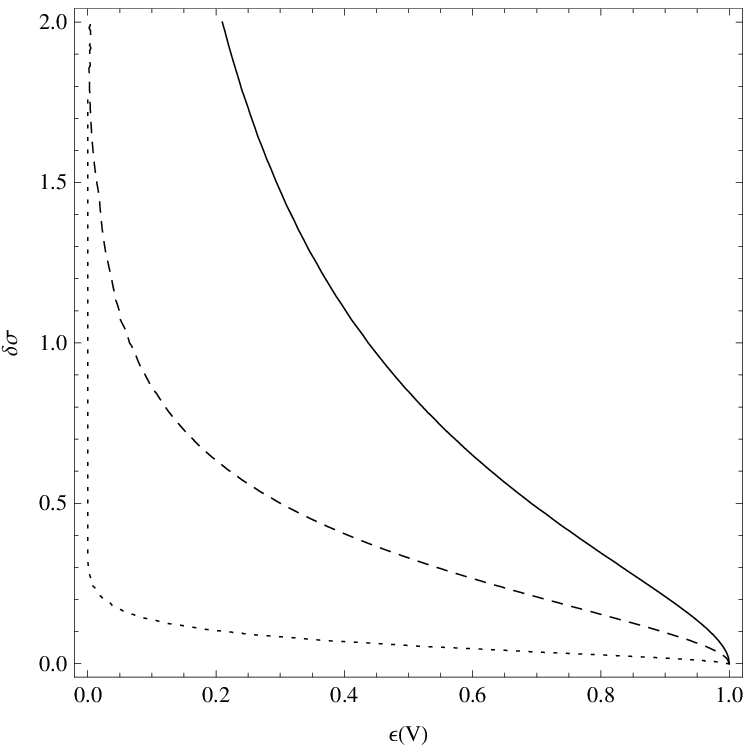, width=6 cm}
\epsfig{file=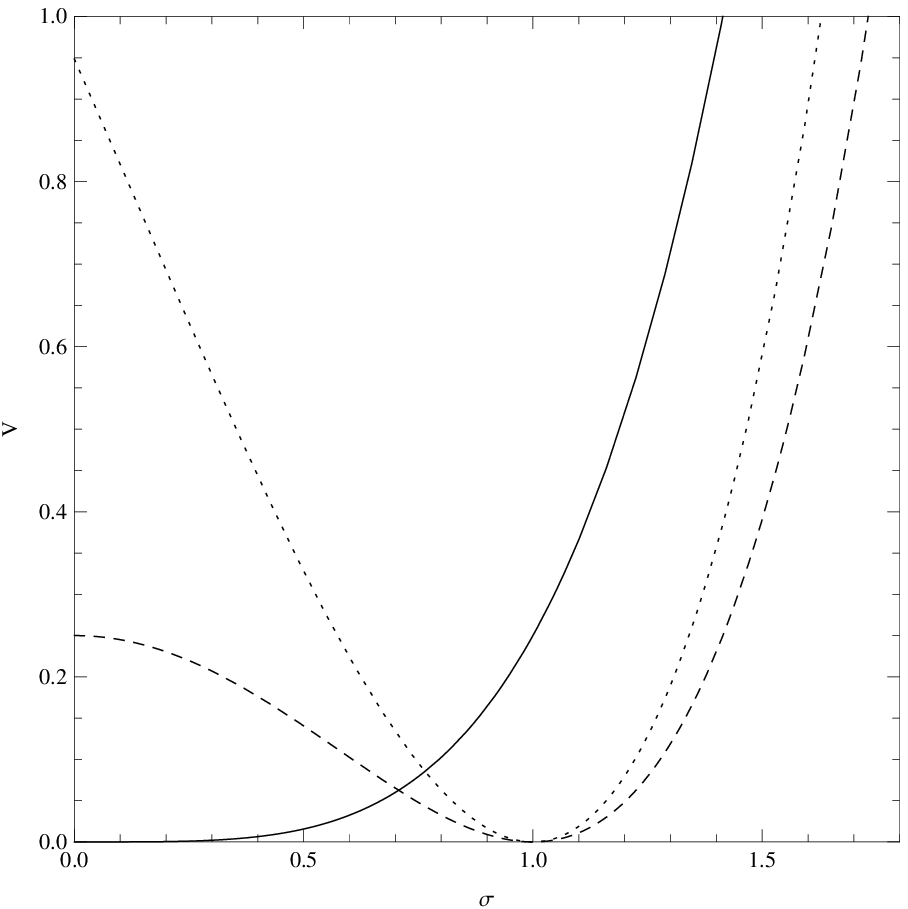, width=6 cm}
\caption{\it In the figure on the left we plotted the ratio (\ref{ratioV}) as a function of $\delta \sigma$ for the LG potential (solid line) and for the modified potential $V_{M}$ with $M$ given by (\ref{Msigma}) with $\delta=1$ (dashed line) and $\delta=10^{-1}$ (dotted line). The modified potential approaches the quartic potential at smaller displacements w.r.t. $V_{LG}$. Smaller values of $\delta$ improve the mechanism. In the figure on the right we compared the shape of the quartic potential (solid line) to the LG potential (dashed line) and the modified potential with $\delta=3$.}.
\label{fig4}
\end{figure}
The mass of the scalar field fluctuations is proportional to $\lambda$ which is small and we thus expect a very light mass. One then needs $\gamma$ small enough to satisfy the PPN constraints. We thus fixed $\gamma$ to its largest value compatible with solar system observations ($\gamma=5 \times 10^{-7}$) and let $\beta$, $\delta$ and $\Lambda$ vary. Smaller values of $\gamma$ lead to similar results.
As in the LG case, $\beta$ and $\delta$ are fixed by minimizing the distance (\ref{defdist}) between $w_{eff}^{\Lambda CDM}$ and $w_{eff}^{M}$. The parameter $\Lambda$ is then determined by the supernovae fit.\\
We determined $\beta_{b}=10^{-12}$ and $\delta_{b}=9.39 \times 10^{-7}$ by first comparing the model with $\Lambda$CDM obtaining the corresponding minimum distance
\be{distBF}
\frac{d(w_{eff}^{M},w_{eff}^{\Lambda CDM})}{||w_{eff}^{\Lambda CDM}||}\simeq 9\times 10^{-6}.
\ee
The two functions $w_{eff}^{M}$ and $w_{eff}^{\Lambda CDM}$ are nearly indistinguishable. Note that $\delta$ is of the same order of magnitude of $\gamma$ and this correspondence holds when varying $\gamma$ for this class of potentials.\\ 
We further find $\Lambda_{b}=9.6\times 10^{-121}$ from SN data corresponding to $\chi^{2}_{m}\simeq167$. It is the same value obtained from the supernovae fit in our fiducial model confirming the fact that the two cosmological dynamics are extremely close. This fact is reinforced by the following results: $\Omega_{DE,0}$ and $\Omega_{M,0}$ are those of $\Lambda CDM$ model within the experimental errors and the same holds for Newton's constant. The time variation of Newton's constant is four orders of magnitude smaller then the larger variation compatible with the observational bounds and the PPN constraints are automatically satisfied by the initial choice of $\gamma$. 
We finally note that the Compton wavelength is now $\lambda_{{\rm Compton}}\simeq 10^{19}\;km$.\\
We end by observing that the choice of (\ref{Msigma}) is one of the many possible choices compatible with the data. The crucial point is that the potential must become ``scale invariant'' as early as possible in order to have a slow-roll so as to minimize the time variation of $G_{N}$ as the field returns to the minimum after the ``kick'' generated by the radiation-matter transition. Such a displacement of the scalar field, driven by the matter content of the Universe, alleviates the coincidence problem and does not prevent the formation of structures for any $\gamma$ compatible with PPN observational bounds. It is clear however that in our model the dark energy is time dependent and will eventually disappear.\\
Let us finally address the possibility of adding an Einstein-Hilbert term to our original action (\ref{lag}) with some gravitational coupling $\tilde {\rm M}_{\rm P}$ such as 
\be{nmcc}
\M^{2}=\tilde {\rm M}_{\rm P}^{2}+\gamma \sigma_{0}^{2}.
\ee 
In this case we have General Relativity plus a non minimally coupled scalar field. Due to (\ref{nmcc}), the coupling between the Ricci scalar and the scalar field in this model is smaller w.r.t. (\ref{lag}) and when a certain amount of matter is added one generally expects a smaller displacement of the scalar field from the minimum. In particular for an Higgs-Inflation model such a coupling is $\frac{\gamma\sigma_{0}^{2}}{\M^{2}}\ll 1$ where $\sigma_{0}$ is the v.e.v. of the Higgs and the displacement is tiny for $\gamma \le 1$. By tuning the self coupling of the LG potential, $\lambda$, one can set the end of the period when the scalar field increases (such a period comes to an end when the ``the pushing force'', proportional to $R$, and the ``pulling force'', which depends on the self-coupling $\lambda$, compensate each other). At that time one may have a scalar field energy density dominating over matter (and consequently a dark energy dominated era) or again a matter dominated era. In the latter case the mechanism does not work. This is exactly what happens for the non minimally coupled inflation-Higgs models. On the other hand, if one considers the case $\frac{\gamma\sigma_{0}^{2}}{\M^{2}}\sim 1$, then dark energy dominates over matter but no major improvements w.r.t. the pure induced gravity case ensues. For this reason, while non minimally coupled models generally give more flexibility in tuning the parameters despite their loss of minimality, we limited our analysis to the pure induced gravity case.
\section*{Conclusions}
In this paper, in induced gravity, we examined the consequences on cosmological evolution of the radiation to matter transition for the case of a scalar field at the end of inflation and reheating when such a field is at the minimum of the scale symmetry breaking potential.
When matter begins to dominate over radiation in the early Universe the scalar field is displaced from the minimum and begins to contribute to the total energy density which drives the cosmological expansion. In the case of a Landau-Ginzburg potential, for a certain range of the parameters $\gamma$ and $\lambda$, the energy density of the scalar field begins to dominate over matter at small redshifts thus generating an accelerated expansion.  The model reproduces a cosmological evolution quite similar to that of the $\Lambda$CDM model. For the LG potential, however, the observational bounds on the value of Newton's constant observed today are not satisfied.\\
Small modifications of the shape of the LG potential in the large field regime, in order to make it then more scale invariant, are indeed sufficient to vastly improve the above result. An example of a modified potential has been proposed and studied. Its dynamics is very close to that of $\Lambda$CDM and the model was also shown to be compatible with the observational constraints on the current value of Newton's constant.\\
We finally observe that, in our model, the dynamics of dark energy, triggered by the matter content of the Universe, alleviates the coincidence problem in the sense that cosmic acceleration is a consequence of, and follows, the radiation to matter transition and the energy density driving such an acceleration must have a value comparable to the matter energy density. Of course in our approach for long enough times the dark energy will disappear as the scalar field slowly returns the potential minimum.

\section*{Acknowledgements}
A.K. was partially supported by the RFBR  grant No 11-02-00643.

\end{document}